\documentclass[10pt,a4paper]{article}

\usepackage[english]{babel}
\usepackage{graphicx}
\usepackage{hyperref} 
\usepackage{geometry}
\usepackage{amsmath}
\usepackage{amssymb}
\usepackage{amsfonts}
\usepackage{multicol}
\usepackage{slashed}
\usepackage{float}
\usepackage{caption}
\usepackage{subcaption}
\usepackage{cite}
\usepackage{amsthm}
\usepackage{xcolor}
\usepackage{musicography}
\numberwithin{equation}{section}

\newcommand{\M}{\mathcal{M}}
\newcommand{\T}{\mathcal{T}}
\newcommand{\R}{\mathbb{R}}
\newcommand{\E}{\mathbb{E}}

\newcommand{\p}{\partial}

\def\XXint#1#2#3{{\setbox0=\hbox{$#1{#2#3}{\int}$ }
		\vcenter{\hbox{$#2#3$ }}\kern-.6\wd0}}

\title{\bf Stochastic Quantization of Relativistic Theories}
\author{Folkert~Kuipers$^1$\thanks{E-mail: F.Kuipers@sussex.ac.uk}\\
	$^1${\em Department of Physics and Astronomy, University of Sussex,}\\{\em Brighton, BN1 9QH, United Kingdom}
}

\begin{document}

\maketitle

\begin{abstract}
It was shown recently that stochastic quantization can be made into a well defined quantization scheme on (pseudo-)Riemannian manifolds using second order differential geometry, which is an extension of the commonly used first order differential geometry. In this letter, we show that restrictions to relativistic theories can be obtained from this theory by imposing a stochastic energy-momentum relation. In the process, we derive non-perturbative quantum corrections to the line element as measured by scalar particles. Furthermore, we extend the framework of stochastic quantization to massless scalar particles.
\end{abstract}

\thispagestyle{empty}
\clearpage
\setcounter{page}{1}

\section{Introduction}
Stochastic quantization is a quantization scheme comparable to canonical quantization and path integral quantization that is employed in the theory of stochastic mechanics~\cite{Fenyes,Kershaw,Nelson:1966sp,NelsonOld,Yasue,Yasue:1981wu,Guerra:1981ie,Guerra:1982fn,Nelson,Zambrini}. Stochastic mechanics is a theory of Newtonian mechanics coupled to a fluctuating Gaussian background field. Due to the coupling to this background field, particles follow stochastic processes instead of deterministic trajectories. The evolution of the probability density of these processes is governed by complex diffusion equations.
\par

Processes described by complex diffusion equations generically have a single well defined position, but two independent well defined velocities. If one imposes there to be a single well defined velocity, one obtains a real diffusion equation that is better known as the heat equation. The process described by the heat equation is the well known dissipative Brownian motion. This dissipative Brownian motion breaks time reversal symmetry.
If, on the other hand, time reversibility is imposed as a constraint, the governing complex diffusion equation is the Schr\"odinger equation. The resulting process is often called a conservative Brownian motion or a Nelson process.
\par

The derivation of the Schr\"odinger equation for a Newtonian system coupled to a time reversible Gaussian background field is the central result of stochastic mechanics. The stochastic quantization scheme that is employed in stochastic mechanics is build upon five fundamental principles: diffeomorphism invariance, gauge invariance, time reversal symmetry, the principle of least action and the background hypothesis.
\par 

The background hypothesis states that all variables in the theory must be promoted to random variables and the trajectories to time reversible semi-martingale processes. The quadratic variation for these stochastic processes is fixed by the background hypothesis. For massive scalar particles the condition on the quadratic variation takes the form\footnote{In order to avoid confusion with the commutator, we denote the quadratic variation with a double bracket $[[X^\mu,X^\nu]]$ instead of a single bracket, which is the more common notation.}
\begin{equation}
	d[[X^\mu,X^\nu]] = \frac{\hbar}{m} h^{\mu\nu} \, d\tau,
\end{equation}
where $h$ is a positive definite metric tensor, obtained from the metric tensor $g$ with Lorentzian signature by a Wick rotation. The construction of this positive definite tensor is discussed in more detail in Ref.~\cite{Dohrn:1985iu} and reviewed in appendix~\ref{ap:BrownMetric}. 
We note that this condition imposes the stochastic part of $X$ to be a scaled Brownian motion by the L\'evy characterization. Furthermore, we remark that this relation is the equivalent of the canonical commutation relation imposed in the canonical quantization scheme.
\par

Stochastic quantization is closely related to the path integral formulation, as it can be regarded as a local construction scheme for path integrals. In imaginary time, this is achieved by the  Feynman-Kac theorem \cite{FKac}, which maps the path integral formulation to the stochastic formulation. Stochastic quantization extends this stochastic formulation to a real time description. An extension of the Feynman-Kac theorem to the real time path integral is given by the Feynman-It\^o theorem \cite{FIto,Albeverio}. Although this theorem does not have an immediate stochastic interpretation, the real time path integral has been related explicitly to the stochastic quantization framework~\cite{Pavon:2000}. 
\par 

The mathematical advantage of the stochastic quantization scheme over the path integral formulation resides in the fact that stochastic integrals are better understood than path integrals. This is an important motivation for the study of stochastic quantization. For similar reasons, the framework is used as an important tool in constructive approaches to quantum field theory \cite{Nelson,Albeverio}. The study of constructive approaches to quantum field theory bears relevance, as the absence of a mathematically rigorous framework of relativistic quantum field theory lies at the heart of several issues in quantum field theory. One of which is the non-renormalizability of gravity as a quantum theory. 
\par

A second motivation for the study of stochastic mechanics is of a foundational nature. The philosophy governing stochastic quantization is closely related to the quantum foam introduced by Wheeler~\cite{Wheeler:1955zz}. However, in stochastic quantization the quantum foam is considered to be the source rather than the consequence of quantum mechanics.
\par 

Stochastic mechanics is a classical\footnote{We call the theory classical, as the quantum configuration space is a covering space over the classical configuration space. The covering is crucial for the treatment of intrinsically quantum properties such as spin and discretized spectra, cf. e.g. Ref.~\cite{Nelson}.} probabilistic\footnote{We call a theory probabilistic, if there is a structure of a probability space $(\Omega,\Sigma,\mathbb{P})$, a measurable configuration space $(\M,\mathcal{B}(\M),\mu)$ and random variables $X:(\Omega,\Sigma,\mathbb{P})\rightarrow (\M,\mathcal{B}(\M))$ such that $\mu=\mathbb{P}\circ X^{-1}$. The random variables are elements of an $L^p$-space. As usual in quantum mechanics, we consider the $L^2$-space, which has the important properties that it is a Hilbert space and that it is self-dual.} interpretation of quantum theory. In this framework, the physical configuration space is a measurable covering space of the classical configuration space. The $L^2$-space containing the wave functions is built on top of this. Although this $L^2$-space is crucial for mathematical analysis, global existence of the wave functions is not required in a stochastic formulation. The wave function represents the best possible prediction of a system given the measurements of the system at earlier times, but is not a physical object. Measuring a system amounts to conditioning the stochastic process.\footnote{We consider measurements where the interaction between the measurement device and the system is negligible. For microscopic systems, such measurements are unachievable. However, these interactions are unrelated to the wave function collapse in the stochastic interpretation.} Collapse of the wave function thus occurs due to updating the filtration to which the process is adapted.\footnote{Let us add a clarification by making a comparison to stock markets: the shares in a stock market have a well defined value at any point in time. However, if we do not observe the value for a certain amount of time, we can only give a probabilistic description of the value of the stock, which is modeled by a probability distribution. Once we decide to observe the market this probability distribution collapses to a delta distribution. According to stochastic mechanics the situation in quantum mechanics is similar. A difference between the two pictures is that quantum mechanics is governed by a time-reversible Brownian motion, while stock markets are usually modeled by a dissipative Brownian motion. As a consequence, quantum mechanics is modeled by a complex wave function, while the probability distributions in stock markets often take the shape of a Gaussian profile. We should stress that the picture is not in conflict with the superposition principle. The superposition principle holds in the stochastic interpretation as particles move between different layers in the covering space. Before measuring a particle, the observers can only give a probabilistic prediction on which layer they will measure the particle, and thus what values of spin or other discretized spectra they will measure. This leads to the superposition principle in the description given by the observer. Furthermore, we emphasize that stochastic mechanics is agnostic about the question whether the quantum fluctuations are fundamental or can be derived from a more fundamental deterministic theory. However, the Bell experiments suggest that the stochasticity is fundamental.}
\par 

Finally, stochastic quantization has received attention, since it can be used as a computational framework in quantum field theory. Stochastic quantization provides an alternative mathematical model that can be used to calculate observables in quantum theories. For certain problems this could simplify the calculations, while other problems are more easily solved using standard quantum field theory methods. Stochastic quantization should therefore be regarded as complementary to other approaches. In this respect, the reformulation due to Parisi and Wu \cite{Parisi:1980ys,Damgaard:1983tq,Damgaard:1987rr} has achieved considerable success in numerical calculations. This reformulation has also been related to quantum gravity inspired theories \cite{Mansi:2009mz,Orlando:2009en,Dijkgraaf:2009gr}.
\par

The general formalism of stochastic quantization is a well defined approach to quantum mechanics for non-relativistic scalar particles on $\R^n$ charged under scalar and vector potentials. Extensions have been made to Riemannian manifolds~\cite{Dankel,DohrnGuerraI,DohrnGuerraII,Dohrn:1985iu,Guerra:1982fn,Nelson}. In addition, particles with spin have been discussed in this framework, cf. e.g. Refs.~\cite{Nelson,Dankel,Fritsche:2009xu}. Furthermore, field theoretic extensions have been developed, see e.g. \cite{Guerra:1973ck,GuerraRuggiero,Guerra:1980sa,Guerra:1981ie,Kodama:2014dba,Marra:1989bi,Morato:1995ty,Garbaczewski:1995fr,Pavon:2001}. We note that the field theoretic framework is more evolved in the  Parisi-Wu formulation. Furthermore, it is worth noticing that many standard quantum mechanics problems have been discussed in the stochastic quantization framework, see e.g. Refs.~\cite{Zambrini,Nelson,Guerra:1981ie,Pena,Olavo,Petroni_2000,Gaeta,Gokler}. Finally, the ideas governing stochastic quantization have been incorporated in models of quantum gravity \cite{Markopoulou:2003ps,Erlich:2018qfc}. For a more complete review of stochastic quantization we refer to Refs.~\cite{Nelson,Guerra:1981ie,Zambrini,Kuipers:2021jlh}.
\par

Most successes of stochastic quantization are of a non-relativistic nature. Although a relativistic version has been treated in the literature, cf. e.g. Refs.~\cite{Guerra:1973ck,GuerraRuggiero,Dohrn:1985iu,Marra:1989bi,Guerra:1981ie,Morato:1995ty,Garbaczewski:1995fr,Pavon:2001}, it is not as well established as the non-relativistic theory. In this letter, we remedy this and show that stochastic quantization can be made into a relativistic quantization scheme. Here, we build on our previous work \cite{Kuipers:2021jlh}, where stochastic quantization was extended to (pseudo-)Riemannian geometry. In this letter, we restrict this general framework to a special class of theories, namely the relativistic theories defined on Lorentzian manifolds. More concretely, we discuss the stochastic quantization of a single relativistic spinless particle on a curved space-time charged under scalar and vector potentials.
\par

A difficulty that arises, when one tries to extend stochastic quantization to (pseudo-)Riemannian manifolds is that there exists a single well defined position $X$, but two independent well defined velocities\footnote{Note that the definition requires to take conditional expectations. Without this conditional expectation there is no notion of velocity, as the stochastic process is almost surely nowhere differentiable.}
\begin{align}\label{eq:Velocity}
	v_+(X(\tau),\tau) &= \lim_{h\downarrow 0}\frac{1}{h} \E\left[X(\tau+h) - X(\tau) | X(\tau)\right],\nonumber\\
	v_-(X(\tau),\tau) &= \lim_{h\downarrow 0} \frac{1}{h} \E\left[X(\tau) - X(\tau-h) | X(\tau)\right],
\end{align}
which are often re-expressed as $v=\frac{1}{2}(v_+ + v_-)$ and $u=\frac{1}{2}(v_+ - v_-)$. These velocity vectors are not vectors in the usual geometrical sense, i.e. they do not transform as vectors under coordinate transformations. Therefore, stochastic quantization cannot be easily embedded in differential geometry, which is the mathematical corner stone of general relativity. This issue was resolved for semi-martingale processes on smooth manifolds with a connection by extending the ordinary first order geometry to a second order geometry, cf. Refs.\cite{Schwartz,Meyer,Emery}. In second order geometry the (co)tangent spaces are extended to second order (co)tangent spaces.  This allows to interpret $v_\pm$ as vectors in these second order spaces. Consequently, the stochastic processes discussed in this paper are diffeomorphism invariant. We refer to Ref.~\cite{Kuipers:2021jlh} for a more detailed exposition of stochastic quantization in the context of second order geometry.
\par 

This paper is organized as follows: in the next section, we discuss relativistic massive theories. In section 3, we extend stochastic quantization to massless theories. In section 4, we discuss the notion of off-shellness in stochastic quantization, and in section 5 we conclude.

\section{Massive scalar particles}
We consider the classical relativistic action
\begin{equation}
	S(x) 
	= 
	- \left[\int m \, \sqrt{- g_{\mu\nu}(x)\, v^\mu\, v^{\nu}}
	+ q\, A_\mu(x)\, v^\mu\right] d\tau
\end{equation}
defined on an $(n=(d+1))$-dimensional Lorentzian manifold $\M$. Following standard procedures we rewrite this action in the form
\begin{equation}
	S(x) 
	= 
	\int \left[
	\frac{e}{2} \left(e^{-2} g_{\mu\nu}(x)\, v^\mu\, v^{\nu}
	- m^2 \right)
	- q\, A_\mu(x)\, v^\mu \right] d\tau,
\end{equation}
where $e$ is an einbein field along the worldline of the particle. As we will consider the equations of motion of massive particles under the gauge fixing condition $e=m^{-1}$, this action is equivalent to the action
\begin{equation}
	S(x) 
	= 
	\int \left[
	\frac{m + \lambda}{2} \, g_{\mu\nu}(x)\, v^\mu\, v^{\nu}
	+ \frac{\lambda}{2} 
	- q\, A_\mu(x)\, v^\mu \right] d\tau,
\end{equation}
where $\lambda$ is a Lagrange multiplier that must be gauge fixed to $\lambda=0$ in the equations of motion. Its equation of motion is algebraic and reproduces the energy-momentum relation
\begin{equation}
	g_{\mu\nu} v^\mu v^\nu = - 1.
\end{equation}
We will thus consider the classical Lagrangian
\begin{equation}\label{eq:LagMassive}
	L_c(x,v) 
	= 
	\frac{m+\lambda}{2}\, g_{\mu\nu}(x)\, v^\mu\, v^{\nu}
	+ \frac{\lambda}{2}
	- q\, A_\mu(x)\, v^\mu.
\end{equation}
If the gauge symmetries of the classical action are to be preserved, the stochastic quantization of this Lagrangian is given by, cf. Ref.~\cite{Zambrini,Kuipers:2021jlh},
\begin{equation}\label{eq:Lagrangian}
	L(X,V,U) 
	= 
	\frac{m+\lambda}{2}\, g_{\mu\nu}(X) \left( V^\mu\, V^{\nu} + U^\mu\, U^{\nu} \right)
	+ \frac{\lambda}{2}
	- q\, A_\mu(X)\, V^\mu,
\end{equation}
where $(X,V,U)$ is a stochastic process on the second order tangent bundle $\hat{T}\M$. $X$ represents the position, $V$ the current velocity and $U$ the osmotic velocity. The corresponding action is given by
\begin{equation}
	S(X) = \E \left[\int L(X,V,U)\, d\tau\right],
\end{equation}
where $\tau$ is the proper time. The equation of motion for $\lambda$ yields the stochastic energy-momentum relation
\begin{equation}\label{eq:StochLineElem1}
	\E \left[ g_{\mu\nu} \left( V^\mu V^\nu + U^\mu U^\nu \right) \right] = - 1
\end{equation}
or equivalently, cf. Ref.~\cite{Kuipers:2021jlh},
\begin{equation}\label{eq:StochLineElem}
	\E \Big[ g_{\mu\nu}\left(dX^\mu dX^\nu + d_\circ \hat{X}^\mu d_\circ \hat{X}^\nu \right) 
	+ \frac{\hbar}{m} \nabla_\mu \left( d_\circ \hat{X}^\mu \right) d\tau - \frac{\hbar^2}{6m^2}\mathcal{R}\, d\tau^2 \Big] = - d\tau^2.
\end{equation}
We note that the geometrical line element remains $g_{\mu\nu}dx^\mu dx^\nu = - d\tau^2$. However, a quantum particle traveling through this geometry does not measure the same length, as it fluctuates around its classical path. Due to these quantum fluctuations, the line element as measured by a quantum particle obtains a quantum correction as given in eq.~\eqref{eq:StochLineElem}. For a single scalar particle adapted to its own natural filtration the osmotic integral vanishes, cf. Ref.~\cite{Kuipers:2021jlh}. This allows to re-express the quantized energy-momentum relation as
\begin{equation}\label{eq:ConditionMass}
	\E \left[ g_{\mu\nu} \, dX^\mu dX^\nu + \left( 1 - \frac{\hbar^2}{6m^2} \mathcal{R} \right)  d\tau^2  \right]
	= 0.
\end{equation}
It follows that scalar quantum particles fluctuate around a quantum corrected path, where the quantum correction is given by the term $\frac{\hbar^2}{6m^2} \mathcal{R}$.
\par 

Minimizing the action leads to the stochastic differential equations in the sense of Stratonovich, cf. Ref.~\cite{Kuipers:2021jlh},
\begin{align}\label{eq:EQM}
	m \, g_{\mu\nu} \left( 
	d^2 X^\nu 
	+ \Gamma^\nu_{\rho\sigma} \, dX^\rho dX^\sigma 
	\right)
	&=
	- \frac{\hbar^2}{12 m} \nabla_\mu \mathcal{R}\, d\tau^2 
	- q \left(
	\nabla_\mu A_\nu - \nabla_\nu A_\mu 
	\right) dX^\nu d\tau
\end{align}
and the condition \eqref{eq:ConditionMass}. When supplemented with the background hypothesis
\begin{equation}\label{eq:BackgroundHypothesis}
	d[[X^\mu,X^\nu]] = \frac{\hbar}{m} h^{\mu\nu}(X) \, d\tau,
\end{equation}
these equations can be solved for the appropriate boundary conditions. The result is a stochastic process $X(\tau)$ parametrized by the proper time. Observables of the theory can be determined from this stochastic process using the standard definitions of the characteristic and moment generating functional
\begin{align}
	\Phi_{X}(J) 
	&= \E\left[ e^{\frac{i}{\hbar} \int J_\mu(\tau) X^\mu(\tau)\, d\tau }\right],\\
	M_{X}(J)
	&= \E\left[ e^{\frac{1}{\hbar} \int J_\mu(\tau) X^\mu(\tau) \, d\tau } \right].
\end{align}
We remark that in contrast to the path integral framework, these expressions do not average over the action. The averaging over the action effectively takes place when the system of equations \eqref{eq:ConditionMass}, \eqref{eq:EQM} and \eqref{eq:BackgroundHypothesis} is solved.
\par

If a probability density $\rho(x,\tau)$ associated to the stochastic process $X$ exists, one can construct the wave function\footnote{Note that the wave function is not always well defined on the configuration space, as this space might not be simply connected. This is the essence of Wallstrom's criticism \cite{Wallstrom:1988zf,WallstromII}. However, if the process is lifted to the universal cover of the configuration space, the wave function $\Psi$ becomes well defined, cf. Ref.~\cite{Nelson}.}
\begin{equation}
	\Psi(x,\tau) = \sqrt{\rho(x,\tau)}\, e^{\frac{i}{\hbar} S(x,\tau)}
\end{equation}
with Hamilton's principal function defined by
\begin{equation}
	S(x,\tau) = \E \left[ \int_{\tau_i}^{\tau} L\big(X,V,U\big) \, dt \Big| X(\tau) = x \right].
\end{equation}
One can then show that this wave function must evolve according to a generalization of the Schr\"odinger equation, cf.~Ref.~\cite{Kuipers:2021jlh} and references therein,
\begin{equation}\label{eq:Schrodinger}
	i \hbar\, \frac{\p}{\p\tau}\Psi 
	=
	- \frac{\hbar^2}{2m} 	\left[ 
	\left(\nabla_\mu + i\, \frac{q}{\hbar}\, A_\mu \right) \left(\nabla^\mu + i\, \frac{q}{\hbar}\, A^\mu \right) 
	- \frac{1}{6} \mathcal{R}
	\right] \Psi.
\end{equation}

As there is no explicit dependence on the affine parameter $\tau$, one can solve eq.~\eqref{eq:Schrodinger} by separation of variables such that
\begin{equation}
	\Psi(x,\tau) = \Phi_\alpha(x)  \exp\left(\frac{i\, m \, \alpha}{2\, \hbar} \tau \right),
\end{equation}
where $\alpha$ is a dimensionless parameter. If we gauge fix $\tau$ to be the proper time, we impose the condition \eqref{eq:StochLineElem1}. Under this constraint the expectation of the energy becomes $-\frac{m}{2}$, which implies $\alpha=1$. We conclude that
\begin{equation}\label{eq:psiSep}
	\Psi(x,\tau) = \Phi(x)  \exp\left(\frac{i\, m}{2\, \hbar} \tau \right),
\end{equation}
where $\Phi(x)$ solves the generalization of the Klein-Gordon equation given by
\begin{equation}
	\left[
	\left(\nabla_\mu + i\,\frac{q}{\hbar}\, A_\mu\right) 
	\left(\nabla^\mu + i\,\frac{q}{\hbar}\, A^\mu\right) 
	- \frac{1}{6} \mathcal{R} 
	- \frac{m^2}{\hbar^2} \right] \Phi 
	= 
	0.
\end{equation}
We remark that the function $\Psi(x,\tau)$ and the relativistic Schr\"odinger equation \eqref{eq:Schrodinger} are not constructed in the traditional approaches to the quantization of relativistic theories. However, their construction is not forbidden in these approaches, while their construction seems necessary in the stochastic approach. The reason for this is that the probability density is defined on the space $\M\times\T$, where $\M$ is the space-time manifold and $\T$ is the proper time monoid. Moreover, the wave function $\Psi(x,\tau)$ is defined on the universal cover\footnote{See previous footnote} of $\M\times\T$.
\par 

An important feature of relativistic theories is that the theory is invariant under proper time reparametrizations. Therefore, one can always perform separation of variables. Consequently, it is sufficient to consider the Klein-Gordon equation for the wave function $\Phi(x)$ in any relativistic quantum theory, as it determines the dynamics of the function $\Psi(x,\tau)$ completely up to a phase factor. This phase factor is given in eq.~\eqref{eq:psiSep} and is a genuine prediction of stochastic quantization.

\section{Massless scalar particles}
Following similar arguments as in previous section using the gauge fixing $e=1$, we obtain the stochastic Lagrangian
\begin{equation}
	L(X,V,U) 
	= 
	\frac{\lambda}{2}\, g_{\mu\nu}(X) \left( V^\mu\, V^{\nu} + U^\mu\, U^{\nu} \right)
	- q\, A_\mu(X)\, V^\mu,
\end{equation}
where the Lagrange multiplier must be gauge fixed to $\lambda=1$ in the equations of motion. The equation of motion for the Lagrange multiplier yields the stochastic energy-momentum relation
\begin{equation}\label{eq:StochLineElemMassless}
	\E \Big[ g_{\mu\nu}\left(V^\mu V^\nu + U^\mu U^\nu \right) \Big] = 0,
\end{equation}
which can be rewritten as
\begin{equation}\label{eq:ConditionMassless}
	\E \left[ g_{\mu\nu} \, dX^\mu dX^\nu - \frac{\hbar^2}{6} \mathcal{R} \,  d\tau^2  \right]
	= 0.
\end{equation}
\par 

Minimizing the action leads to stochastic differential equations in the sense of Stratonovich given by
\begin{align}\label{eq:EQMMassless}
	g_{\mu\nu} \left( 
	d^2 X^\nu 
	+ \Gamma^\nu_{\rho\sigma} \, dX^\rho dX^\sigma 
	\right)
	&=
	- \frac{\hbar^2}{12 m} \nabla_\mu \mathcal{R} \, d\eta^2 
	- q \left(
	\nabla_\mu A_\nu - \nabla_\nu A_\mu 
	\right) dX^\nu d\eta
\end{align}
and the constraint \eqref{eq:ConditionMassless}. We note that $\eta$ is an affine parameter that has the dimension of time per unit mass. The background hypothesis in the massless case under the gauge fixing $\lambda=1$ takes the shape
\begin{equation}\label{eq:BackgroundHypothesisMassless}
	d[[X^\mu,X^\nu]] = \hbar \, h^{\mu\nu}(X) \, d\eta.
\end{equation}
The system of equations \eqref{eq:ConditionMassless}, \eqref{eq:EQMMassless} and \eqref{eq:BackgroundHypothesisMassless} can be solved for the appropriate boundary conditions. The result is a stochastic process $X(\eta)$ parametrized by the parameter $\eta$. Observables of the theory can be determined from this stochastic process using the characteristic and moment generating functional.
\par 

The derivation of the Schr\"odinger equation in the massless case is similar to the derivation in the massive case, which can be found in Ref.~\cite{Kuipers:2021jlh} and references therein. If a probability density $\rho(x,\eta)$ associated to the stochastic process $X$ exists, one can construct the wave function
\begin{equation}
	\Psi(x,\eta) = \sqrt{\rho(x,\eta)} \, \exp \left\{ \frac{i}{\hbar}\, \E \left[ \int_{\eta_i}^{\eta} L\big(X(t),V(t),U(t),t\big) \, dt \Big| X(\eta) = x \right] \right\}	
\end{equation}
that evolves according to a generalization of the Schr\"odinger equation
\begin{equation}
	i \hbar\, \frac{\p}{\p\eta}\Psi 
	=
	- \frac{\hbar^2}{2}\left[
	\left(\nabla_\mu + i\,\frac{q}{\hbar}\, A_\mu \right) 
	\left(\nabla^\mu + i\,\frac{q}{\hbar}\, A^\mu \right) 
	- \frac{1}{6} \mathcal{R} \right] \Psi.
\end{equation}
As there is no explicit dependence on the affine parameter $\eta$, one can solve by separation of variables, such that
\begin{equation}
	\Psi(x,\eta) = \Phi_\alpha(x)  \exp\left(\frac{ i \, \hbar \, \alpha}{2} \, \eta \right),
\end{equation}
where $\alpha$ has the dimension of inverse length squared. If we impose the condition \eqref{eq:StochLineElemMassless}, the kinetic energy becomes $0$. This imposes $\alpha=0$. We conclude that
\begin{equation}\label{eq:PsiMassless}
	\Psi(x,\eta) = \Phi(x),
\end{equation}
where $\Phi(x)$ solves the generalization of the Klein-Gordon equation given by
\begin{equation}
	\left(
	\Big[\nabla_\mu + i\,\frac{q}{\hbar}\, A_\mu\Big] 
	\Big[\nabla^\mu + i\,\frac{q}{\hbar}\, A^\mu\Big] 
	- \frac{1}{6} \mathcal{R} 
	\right) \Phi 
	= 
	0.
\end{equation}
We remark that the vanishing phase factor in eq.~\eqref{eq:PsiMassless} is expected, as massless particles are restricted to $d$-dimensional submanifolds of $\M$.

\section{Off-Shell Motion}
Let us consider the Lagrangian \eqref{eq:LagMassive} for a massive particle in the simple case that $g_{\mu\nu}=\eta_{\mu\nu}$ and $q=0$. The on-shell condition \eqref{eq:ConditionMass} is then given by
\begin{equation}
	\E \Big[ \eta_{\mu\nu} \, d X^\mu d X^\nu  \Big] = - d\tau^2.
\end{equation}
Moreover,
\begin{align}
	d X^\mu(\tau) = v^\mu(\tau)\, d\tau + \frac{1}{2} \Big( d W_+^\mu(\tau) + d W_-^\mu(\tau) \Big),
\end{align}
where $W_\pm$ are independent Brownian motions, cf. e.g. Ref.~\cite{Kuipers:2021jlh} and references therein.
Consequently,
\begin{align}
	\E \Big[ \eta_{\mu\nu} \, d X^\mu d X^\nu  \Big] 
	&= 
	\E\Big[\eta_{\mu\nu} \Big(
	v^\mu v^\nu d\tau^2
	+ v^\mu \big( d W_+^\nu + d W_-^\nu \big) d\tau \nonumber\\
	&\qquad \quad \left. \left.
	+ \frac{1}{4} \big( d W_+^\mu(\tau) + d W_-^\mu(\tau) \big) \big( d W_+^\nu + d W_-^\nu \big) \right) \right]\nonumber\\
	&=
	\eta_{\mu\nu} \, v^\mu v^\nu\, d\tau^2,
\end{align}
where we used that
\begin{align}
	\E\left[d W_\pm^\mu \right] &= 0,\\
	\E\left[d W_+^\mu d W_-^\nu \right] &= \E\left[d W_+^\mu\right] \E\left[ d W_-^\nu \right],\\
	\E\left[ d W_+^\mu d W_+^\nu \right] &= - \E\left[ d W_-^\mu d W_-^\nu\right].
\end{align}
The first equation follows from the fact that $W_\pm$ is a martingale, the second from the stochastic independence of the forward and backward processes $W_+$ and $W_-$, and the third from the time reversibility of the semi-martingale $X$.
\par

Under the expectation value the particle moves on-shell, i.e. 
\begin{equation}
	\eta_{\mu\nu} \, v^\mu v^\nu = -1.
\end{equation}
However, without the expectation value this relation is not satisfied. Therefore, the expected trajectory of a particle is on-shell, but the actual trajectory of a particle can be off-shell.
As $dW_\pm^\mu(\tau) \sim \mathcal{N}\left(0,\frac{\hbar}{m}d\tau\right)$, it is easy to see that the quantum fluctuations dominate in the regime
\begin{equation}
	c \, d\tau \lesssim \frac{\hbar}{m\, c},
\end{equation}
which corresponds to length scales less than the de Broglie wavelength. On these length scales the event $\{\eta_{\mu\nu} \, d X^\mu d X^\nu\geq0\}$ becomes very likely. Therefore, according to stochastic mechanics, particles have a high probability of traveling faster than light on length scales less than the de Broglie wavelength, while the probability of traveling faster than light over length scales larger than the de Broglie wavelength quickly decays to $0$. According to the stochastic interpretation, this is the reason why particles are not localized within their de Broglie wavelength. We remark that this interpretation is given in a position representation, where the process $(X,V,U)$ is adapted to the natural filtration of $X$. In other words we perform position measurements only. As is the case in other quantization schemes, stochastic quantization predicts an uncertainty relation between position and momentum measurements.
\par

We note that this result is similar in the path integral approach. However, there is a difference in the interpretation: in the stochastic approach there is a single well defined stochastic trajectory, while the path integral approach considers the statistical ensemble of the sample paths of the stochastic trajectory. These sample paths are virtual and in this approach there is no notion of the real trajectory. From the perspective of modern probability theory, the path integral can in principle be derived from the stochastic integral, if both are well defined. Consequently, it is unlikely that the two interpretations can be distinguished experimentally, as their physical predictions are equivalent.

\section{Conclusion}
In this letter, we have shown that stochastic quantization can be made into a well defined quantization scheme for relativistic theories. Furthermore, we have extended the framework such that it includes massless particles. We point out that stochastic quantization is a local quantization scheme and that the motion of particles in this framework is governed by stochastic differential equations. In this framework, the Schr\"odinger equation and Klein-Gordon equation are derived from first principles. Finally, we have discussed the interpretation of off-shellness in the stochastic framework. We conclude that stochastic quantization is an interesting framework with important implications for the mathematical and philosophical foundations of quantum theory.

\section*{Acknowledgments}
This work is supported by a doctoral studentship of the Science and Technology Facilities Council. I am grateful to Xavier Calmet for helpful comments on the manuscript.

\appendix

\section{Construction of the Brownian metric}\label{ap:BrownMetric}
The background field hypothesis was introduced in Ref.~\cite{Nelson} for massive particles as
\begin{equation}
	d[[X^\mu,X^\nu]] = \frac{\hbar}{m} g^{\mu\nu}(X) \, d\tau.
\end{equation}
If $g^{\mu\nu}$ has a definite signature this condition has a semi-martingale solution, but for indefinite signature there exist no semi-martingale satisfying this condition. The extension of this condition to manifolds with indefinite signatures, and in particular with a Lorentzian signature has been the subject of several studies, see e.g. Refs.~\cite{Guerra:1973ck,GuerraRuggiero,Dohrn:1985iu,Marra:1989bi,Guerra:1981ie,Morato:1995ty,Garbaczewski:1995fr,Pavon:2001}. In this paper, we adopt the approach discussed in Ref.~\cite{Dohrn:1985iu}.
\par 

We reformulate the background hypothesis as
\begin{equation}
	d[[X^\mu,X^\nu]] = \frac{\hbar}{m} h^{\mu\nu}(X) \, d\tau,
\end{equation}
where $h^{\mu\nu}$ is a positive definite tensor that is sometimes called the \textit{Brownian metric}. Its inverse $h_{\mu\nu}$ is defined by the relation 
\begin{equation}
	h_{\mu\nu} h^{\nu\rho} = \delta_\mu^\rho.
\end{equation}
Moreover, it is related to the kinetic metric $g_{\mu\nu}$ through the compatibility condition
\begin{equation}
	g_{\mu\nu} h^{\mu\rho} h^{\nu\sigma} = g^{\rho\sigma}.
\end{equation}
where $g^{\mu\nu}$ is the inverse of the \textit{kinetic metric} $g_{\mu\nu}$. If the kinetic metric $g_{\mu\nu}$ has a definite signature, the compatibility condition yields a unique solution for the Brownian metric $g^{\mu\nu}=h^{\mu\nu}$, but for a Lorentzian signature there is a family of positive definite solutions $h^{\mu\nu}$. In this paper, we work in the $(-+...+)$ convention and set
\begin{equation}
	h^{\mu\nu} = g^{\mu\nu} + 2\, u^\mu u^\nu
\end{equation}
with time-like vector $u^\mu=(1,0,...,0)$, which is uniquely defined and satisfies the given conditions.
\par 

We remark that in order to obtain a covariant stochastic theory, we have adopted the Schwartz-Meyer second order geometry framework discussed in Refs.~\cite{Schwartz,Meyer,Emery}. In a local coordinate system second order vectors can be expressed as
\begin{equation}
	V = v^\mu \, \p_\mu + v^{\mu\nu} \, \p_\mu \p _\nu,
\end{equation}
where $v^{\mu\nu}$ is the second order part of the vector $v$. As discussed in Ref.~\cite{Kuipers:2021jlh}, the background hypothesis fixes the second order part of the velocity vectors such that
\begin{equation}
	v_\pm^{\mu\nu} = \pm \frac{\hbar}{2m} \,g^{\mu\nu}\, d\tau
\end{equation}
which is defined in terms of the kinetic metric. Consequently, the kinetic equations \eqref{eq:ConditionMass}, \eqref{eq:EQM} and \eqref{eq:Schrodinger} are independent of the Brownian metric, as was already observed in Ref.~\cite{Dohrn:1985iu}.
\par 

Finally, we notice that the constructions in this appendix can be generalized straightforwardly to the massless case discussed in section 3.

\end{document}